\documentstyle[12pt]{article}
\begin{document}
\tolerance=5000
\def\nn{\nonumber \\}
\def\be{\begin{equation}}
\def\ee{\end{equation}}
\def\bea{\begin{eqnarray}}
\def\eea{\end{eqnarray}}
\def\tr{{\rm tr}}
\def\cL{{\cal L}}
\def\e{{\rm e}}
\def\eg{{\it e.g.\ }}
\def\ie{{\it i.e.},\ }
\def\etc{{\it etc}.\ }

\begin{flushright}
\begin{minipage}{3cm}
OCHA-PP-93 \\
NDA-FP-31 \\
March (1997)
\end{minipage} 
\end{flushright}
\vfill
\begin{center}
{\large\bf $N=3$ and $N=4$ Two-Form 
Supergravities}

\vfill

{\sc Tomoko KADOYOSHI}\footnote{
e-mail : kado@fs.cc.ocha.ac.jp} and 
{\sc Shin'ichi NOJIRI}$^\spadesuit$\footnote{
e-mail : nojiri@cc.nda.ac.jp}

\vfill

{\it Department of Physics, Faculty of Science}

{\it Ochanomizu University}

{\it Otsuka, Bunkyo-ku, Tokyo 112, JAPAN}

and

$\ ^\spadesuit$ {\it Department of Mathematics and 
Physics}

{\it National Defence Academy}

{\it Hashirimizu, Yokosuka 139, JAPAN}

\vfill

{\bf ABSTRACT}
\end{center}

We construct the Lagrangeans of $N=3$ and $N=4$ two-form 
supergravities. The two-form gravity theories are classically 
equivalent to the Einstein gravity theories and 
can be formulated as gauge theories. 
The gauge algebras used here can be identified 
with the subalgebra of $N=3$ superconformal algebra and 
$SU(2)\times SU(2)\times U(1)$-extended $N=4$ superconformal 
algebra.

\newpage

It should be natural to regard the Einstein gravity theory 
as an effective theory of 
a more fundamental theory \eg superstring theory
since the Lagrangean of the Einstein gravity is not 
renormalizable.
Two-form gravity theory is classically 
equivalent to the Einstein gravity theory and 
appeared as the Lagrangean formalism of the 
Ashtekar formulation \cite{astk}.
The two-form gravity is obtained 
from the BF theory \cite{schw}, which is 
a topological field theory, 
by imposing constraint conditions \cite{pl}.
The BF theory has a large local symmetry called 
the Kalb-Ramond symmetry \cite{kara}.
Since the Kalb-Ramond symmetry is very stringy symmetry, 
the fundamental gravity theory is expected to be a kind of 
string theory \cite{kkns}.

The two-form supergravity theory was considered in 
Ref.\cite{jacob} and the two-form supergravity theories 
which have a cosmological term and $N=2$ supersymmetry 
were proposed in Ref.\cite{ks}.
The group theoretical structure of 
these supergravity theories was discussed 
in Refs.\cite{ezawa,nojiri}
and it was clarified that $N=1$ and $N=2$ two-form 
supergravity theories can be formulated as gauge theories.
The gauge algebras are the subalgebras of $N=1$ and $N=2$ 
Neveu-Schwarz algebaras whose generators are 
$(L_0, L_{\pm 1}, G_{\pm{1 \over 2}})$ and 
$(L_0, L_{\pm 1}, G^\pm_{\pm{1 \over 2}}, J_0)$, 
respectively.
Then it might be natural to expect that $N=3$ and $N=4$ 
two-form supergravities can be formulated as the gauge 
theories whose gauge algebras are the subalgebras of $N=3$ 
and $N=4$ \cite{ss,stvp} 
Neveu-Schwarz algebra.
In this paper, we construct the Lagrangean 
of $N=3$ and $N=4$ two-form supergravity
based on the subalgebra of 
$SU(2)\times SU(2)\times U(1)$-extended $N=4$ 
superconformal (Neveu-Schwarz) algebra \cite{stvp}, 
which is known 
to be an irreducible algebra containing the largest 
number of supercurrents.

The $SU(2)\times SU(2)\times U(1)$-extended $N=4$ 
superconformal (Neveu-Schwarz) algebra is composed of 
Virasoro operators, four supercurrents, two $SU(2)$ 
currents, four free fermion currents and $U(1)$ current 
(free boson).
The zero modes of two $SU(2)$ currents,  
$\pm {1 \over 2}$-modes of four supercurrents 
and $0$,$\pm 1$-modes of Virasoro generators make a 
closed subalgebra, which we call $N=4$ topological 
superalgebra ($N=4$ TSA) and can be expressed as follows
\bea
\label{n4al}
&& [A^a, A^b]=i\epsilon^{abc}A^c\ , \hskip 0.5cm
[B^a, B^b]=i\epsilon^{abc}B^c\ , \hskip 0.5cm
[T^a, T^b]=i\epsilon^{abc}T^c\ ,\nn
&& [A^a, B^b]=[B^a, T^b]=[T^a, A^b]=0 \ , \nn
&& [G_{KL\, A}, A^a]=T^{a\ M}_K G_{ML\, A}\ , \hskip 1cm
[G_{KL\, A}, B^a]=T^{a\ M}_L G_{KM\, A}\ , \nn
&& [G_{KL\, A}, T^a]=T^{a\ B}_A G_{KL\, B}\ , \nn
&& \{G_{KL\, A}, G_{MN\, B}\}=\alpha T^a_{KM}
\epsilon_{LN}\epsilon_{AB}A^a
+\beta \epsilon_{KM} T^a_{LN} \epsilon_{AB} B^a \nn
&& \ \hskip 4cm +\epsilon_{KM} \epsilon_{LN} T^a_{AB} T^a \ .
\eea
Here $A,B,\cdots , K,L, \cdots=1,2$\footnote{
We use $A,B,\cdots$ as spinor indices with respect to 
$SU(2)$ generated by $T^a$ and $K,L,\cdots$ with respect 
to two $SU(2)$'s generated by $A^a$ and $B^a$.
}, $a,b,\cdots =1,2,3$, 
\be
\label{T}
T^{a\ B}_{\ A}={1 \over 2}\sigma^{a\ B}_{\ A} 
\ee
($\sigma^a$'s ($a=1,2,3$) are the Pauli matrices.)
\bea
\label{T2}
T^a_{AB}&\equiv& \epsilon_{BC}T^{a\ C}_{\ A} \nn
T^{aAB}&\equiv& \epsilon^{AC}T^{a\ B}_{\ C} 
\eea
and 
\bea
\label{eps}
&& \epsilon^{AB}=-\epsilon^{BA} \nn
&& \epsilon_{AB}=-\epsilon_{BA} \nn
&& \epsilon^{12}=\epsilon_{21}=1 \ .
\eea
In the above algebra (\ref{n4al}),  $A^a$ and $B^a$ 
correspond to the zero modes of 
two $SU(2)$ currents, $G_{KL\,A}$ to 
$\pm {1 \over 2}$-modes of four 
supercurrents and  
$T^a$ to $0$,$\pm 1$-modes of Virasoro operators 
in the $SU(2)\times SU(2)\times U(1)$-extended $N=4$ 
superconformal (Neveu-Schwarz) algebra.
The Jacobi identity
\bea
\label{jacobi}
&& [G_{KL\, A}, \{G_{MN\, B},G_{PQ\, C}\}]
+[G_{MN\, B}, \{G_{PQ\, C}, G_{KL\, A}\}] \nn
&& \ \hskip 2cm+[G_{PQ\, C},\{G_{KL\, A}, G_{MN\, B}\}]=0 
\eea
requires
\be
\label{ab1}
\alpha+\beta+1=0\ .
\ee
The invariant traces of the product of the two operators 
can be defined, up to a multiplicative constant, to be
\bea
\label{tr4}
&& \tr T^aT^b = \delta^{ab}\ ,\hskip 1cm 
\tr A^aA^b={1 \over \alpha}\delta^{ab}\ ,\hskip 1cm 
\tr B^aB^b={1 \over \beta}\delta^{ab} \nn
&& \tr G_{KL\,A} G_{MN\,B}=-\epsilon_{KM}\epsilon_{LN}
\epsilon_{AB}\ .
\eea

The $N=4$ TSA contains an algebra which has three 
supercurrents as a subalgebra. 
We call the algebra as $N=3$ topological superalgebra 
($N=3$ TSA).
The reduction to $N=3$ TSA from $N=4$ TSA is given by 
\bea
\label{red}
&& G^a_A \equiv T^{aKL}G_{KL\, A}\ , \hskip 1cm 
J^a\equiv A^a+B^a\ , \nn
&& \alpha=\beta=-{1 \over 2} \ .
\eea
The $N=3$ TSA is 
\bea
\label{al3}
&& [J^a, J^b]=i\epsilon^{abc}J^c\ , \hskip 1cm
[T^a, T^b]=i\epsilon^{abc}T^c\ ,\nn
&& [J^a, G^b_A]=i\epsilon^{abc}G^c_A\ , \hskip 1cm
[G^a_A, T^b]=T^{a\ B}_{\ A}G^b_B \ ,\nn
&& \{G^a_A, G^b_B\}={i \over 8}\epsilon^{abc}\epsilon_{AB}J^c
-{1 \over 2}\delta^{ab}T^c_{AB}T^c
\eea
and the invariant traces are
\be
\label{tr3}
\tr T^aT^b = \delta^{ab}\ ,\hskip 1cm \tr J^aJ^b=
-4\delta^{ab}\ ,\hskip 1cm 
\tr G^a_A G^b_B={1 \over 2}\delta^{ab}\epsilon_{AB}\ .
\ee
As we will see later, $G_{KL\,A}$ and $G^a_A$ generate 
left-handed supersymmetry.

Two-form gravity theory is obtained 
from the BF theory by imposing constraint conditions \cite{pl}.
In order to keep the symmetries given by $N=3$ or $N=4$ TSA 
when imposing the constraints, 
we introduce the multiplier fields later. 
The $N=3$ or $N=4$ TSA multiplet of 
the multiplier field needs to 
contain a field of spin 2 representation with respect 
to $SU(2)$ generated by $T^a$. 
The representations containing the higher spins can 
be constructed by using a tensor product, where  
$T^a$ \etc are replaced 
by $T^a \otimes 1 + 1 \otimes T^a$ \etc \cite{nojiri}.
In case of $N=4$ TSA, such a multiplet is given by 
\bea
\label{KN4}
M_{ABCD}&=&T^a_{(AB}T^b_{CD)}T^a\otimes T^b \ , \nn
N_{KL\,ABC}&=&T^a_{(AB}(G_{KL\,C)}\otimes T^a
+T^a \otimes G_{KL\,C)}) \ , \nn
K^{ab}(\alpha)&=&\alpha(A^a\otimes T^b+T^b\otimes A^a)
+2T^{aKM}\epsilon^{LN}T^{bAB}G_{KL\, A}\otimes 
G_{MN\, B} \ , \nn
K^{ab}(\beta)&=&\beta(B^a\otimes T^b+T^b\otimes B^a) 
+2\epsilon^{KM}T^{aLN}T^{bAB}G_{KL\, A}\otimes 
G_{MN\, B} \ , \nn
L_{KL\,A}&=&{2 \over 3}(\alpha-\beta)
T^{a\ B}_{\ A}(G_{KL\,B}\otimes T^a +T^a \otimes G_{KL\,B})
\nn
&& +2\alpha T^{a\ M}_{\ K}
(G_{ML\,A}\otimes A^a +A^a \otimes G_{ML\,A}) \nn
&& -2\beta T^{a\ N}_{\ L}
(G_{KN\,A}\otimes B^a +B^a \otimes G_{KN\, A}) \ , \nn
H&=&{1 \over 3}(\alpha -\beta)T^a\otimes T^a 
+\alpha^2 A^a\otimes A^a - \beta^2 B^a \otimes B^a \nn
&& +{1 \over 2}(\alpha -\beta)\epsilon^{KM}\epsilon^{LN}
\epsilon^{AB}G_{KL\,A}G_{MN\,B} \ .
\eea
Here $(AB\cdots D)$ means a symmetrization with respect 
to $AB\cdots D$  which are indices of 
$SU(2)$ generated by $T^a$.
The multiplet in $N=3$ TSA is also given by,
\bea
\label{KN3}
M_{ABCD}&=&T^a_{(AB}T^b_{CD)}T^a\otimes T^b \ , \nn
N^a_{ABC}&=&T^b_{(AB}(G^a_{C)}\otimes T^b
+T^b \otimes G^a_{C)}) \ , \nn
K^{ab}&=&{1 \over 2}(J^a\otimes T^b+T^b\otimes J^a)
+4i\epsilon^{acd}T^{bAB}G^c_A\otimes G^d_B \ , \nn
L_A&=&(G^a_A\otimes J^a +J^a \otimes G^a_A) \ .
\eea
$(M_{ABCD}$, $N_{KL\,ABC}$, $K^{ab}(\alpha)$, $K^{ab}(\beta)$, 
$L_{KL\,A}$, $H)$ makes a representation of $N=4$ TSA.
The (anti-) commutators of $G_{KL\,A}$ with 
$(M_{ABCD}$, $N_{KL\,ABC}$, $K^{ab}(\alpha)$, $K^{ab}(\beta)$, 
$L_{KL\,A}$, $H)$  are given by\footnote{
$G_{KL\,A}$ in Equation (\ref{KNNM}) 
is understood to be $G_{KL\,A}\otimes 1+1\otimes G_{KL\,A}$.}
\bea
\label{KNNM}
\left[ G_{KL\,A},M_{BCDE} \right] &=&
-{1 \over 2}\epsilon_{A(B} N_{KL\,CDE)} \ , \nn
\{G_{KL\,A}, N_{MN\,BCD}\}&=&
-T^a_{KM}\epsilon_{LN}\epsilon_{A(B}T^b_{CD)}K^{ab}(\alpha) \nn 
&&-\epsilon_{KM}T^a_{LN}\epsilon_{A(B}T^b_{CD)}K^{ab}
(\beta) 
+2\epsilon_{KM}\epsilon_{LN}M_{ABCD} \ , \nn 
\left[ G_{KL\,A}, 
K^{ab}(\alpha) \right] &=&T^{a\ M}_K T^{b\ B}_A
L_{ML\,B}-2(\alpha-1)T^{a\ M}_K T^{aBC}N_{ML\,ABC} \ , \nn
\left[ G_{KL\,A}, K^{ab}(\beta) \right] &=&-T^{a\ N}_L T^{b\ B}_A
L_{KN\,B}-2(\beta-1)T^{a\ N}_L T^{aBC}N_{KN\,ABC} \ , \nn
\{G_{KL\,A}, L_{MN\,B}\}&=&\epsilon_{KM}\epsilon_{LN}
\epsilon_{AB}H 
+{4 \over 3}(\alpha + 2)T^a_{KM}\epsilon_{LN}T^b_{AB}
K^{ab}(\alpha) \nn
&&-{4 \over 3}(\beta + 2)\epsilon_{KM}T^a_{LN}T^b_{AB}
K^{ab}(\beta) \ , \nn
\left[ G_{KL\,A}, H \right]&=&-{1 \over 4}L_{KL\,A} \ .
\eea
In case of $N=3$ TSA, the (anti-)commutaters of $G^a_A$ 
with $(M_{ABCD}$, $N^a_{ABC}$, $K^{ab}$, $L_A)$ are 
given by
\bea
\label{KNNM3}
\left[ G^a_A, M_{BCDE} \right]&=&
-{1 \over 2}\epsilon_{A(B} N^a_{CDE)}
\ , \nn
\{G^a_A, N^b_{BCD}\}&=&
-{i \over 4}\epsilon^{abc}\epsilon_{A(B}T^d_{CD)}K^{cd} 
-{1 \over 2}\delta^{ab}M_{ABCD} \ , \nn
\left[ G^a_A, K^{bc} \right]&=&
-{1 \over 2}\delta^{ab} T^{c\ B}_A
L_B+3i\epsilon^{abd} T^{cBC}N^d_{ABC} \ , \nn
\{G^a_A, L_B\}&=&-T^b_{AB}K^{ab} \ .
\eea
In the following, we call the representations given by 
$(M_{ABCD}$, $N_{KL\,ABC}$, $K^{ab}(\alpha)$, $K^{ab}(\beta)$, 
$L_{KL\,A}$, $H)$ or 
$(M_{ABCD}$, $N^a_{ABC}$, $K^{ab}$, $L_A)$ 
as $N=4$ or $N=3$ spin 2 representation since the highest 
spin with respect to $T^a$ in the representation is 2.

In order to construct the Lagrangeans of $N=3$ and 
$N=4$ two-form supergravity theories, we introduce 
the gauge field $V_\mu$.
In case of $N=4$ TSA, the gauge field is defined by
\be
\label{gauge4}
V_\mu = A^a_\mu A^a + B^a_\mu B^a 
+\psi^{KL\, A}_\mu G_{KL\,A} + \omega^a_\mu T^a \ .
\ee
$N=3$ TSA gauge field is also defined by 
\be
\label{gauge3}
V_\mu = A^a_\mu J^a  
+\psi^{aA}_\mu G^a_A + \omega^a_\mu T^a \ .
\ee
$\omega^a_\mu$ is identified with left-handed spin 
connection and $\psi^{ABC}_\mu$ and $\psi^{aA}_\mu$ are 
with the left-handed Rarita-Schwinger fields.
The field strength (gauge curvature) 
\be
\label{gacu}
R_{\mu\nu}=[\partial_\mu+V_\mu,\partial_\nu+V_\nu]
\ee
has the following form in case of $N=4$ TSA
\bea
\label{gacu4}
R_{\mu\nu}&=&\{\partial_\mu A^a_\nu-\partial_\nu A^a_\mu
+i\epsilon^{abc}A^b_\mu A^c_\nu 
-\alpha T^a_{KM}\epsilon_{LN}\epsilon_{AB}
\psi^{KL\,A}_\mu\psi^{MN\,B}_\nu\}A^a \nn
&& +\{\partial_\mu B^a_\nu-\partial_\nu B^a_\mu
+i\epsilon^{abc}B^b_\mu B^c_\nu 
-\beta \epsilon_{KM}T^a_{LN}\epsilon_{AB}
\psi^{KL\,A}_\mu\psi^{MN\,B}_\nu\}B^a \nn
&& +\{\partial_\mu\psi^{KL\,A}_\nu
-\partial_\nu\psi^{KL\,A}_\mu
+T^{a\ A}_B(\omega_\mu^a \psi_\nu^{KL\,B}
-\omega_\nu^a\psi_\mu^{KL\,B}) \nn
&&\ \ +T^{a\ K}_M(A_\mu^a \psi_\nu^{ML\,A}
-A_\nu^a\psi_\mu^{ML\,A}) \nn
&&\ \ +T^{a\ L}_N(B^a \psi_\nu^{KN\,A}
-B_\nu^a\psi_\mu^{KN\,A})\} G_{KL\,A} \nn
&& +\{\partial_\mu\omega_\nu^a-\partial_\nu\omega_\mu^a
+i\epsilon^{abc}\omega_\mu^b\omega_\nu^c 
-\epsilon_{KM}\epsilon_{LN}T^a_{AB}
\psi^{KL\,A}_\mu\psi^{MN\,B}_\nu\}T^a \ , 
\eea
and $N=3$ TSA
\bea
\label{gacu3}
R_{\mu\nu}&=&\{\partial_\mu A^a_\nu-\partial_\nu A^a_\mu
+i\epsilon^{abc}A^b_\mu A^c_\nu 
-{i \over 8}i\epsilon^{abc}\epsilon_{AB}
\psi^{b\,A}_\mu\psi^{cB}_\nu\}J^a \nn
&& +\{\partial_\mu\psi^{a\,A}_\nu
-\partial_\nu\psi^{a\,A}_\mu
+T^{b\ A}_B(\omega_\mu^b \psi_\nu^{a\,B}
-\omega_\nu^b\psi_\mu^{a\,B}) \nn
&&\ \ +i\epsilon^{abc}(A_\mu^b \psi_\nu^{c\,A}
-A_\nu^b\psi_\mu^{c\,A})\} G^a_A \nn
&&+\{\partial_\mu\omega_\nu^a-\partial_\nu\omega_\mu^a
+i\epsilon^{abc}\omega_\mu^b\omega_\nu^c 
+{1 \over 2}T^a_{AB}
\psi^{b\,A}_\mu\psi^{b\,B}_\nu\}T^a \ .
\eea

We also introduce the two-form field $X_{\mu\nu}$; 
in case of $N=4$ TSA
\be
\label{2f4}
X_{\mu\nu} = \Theta^a_{\mu\nu} A^a + \Pi^a_{\mu\nu} B^a 
+\chi^{KL\, A}_{\mu\nu} G_{KL\,A} + \Sigma^a_{\mu\nu} T^a
\ee
and in case of $N=3$ TSA
\be
\label{2f3}
X_{\mu\nu} = \Pi^a_{\mu\nu} J^a  
+\chi^{aA}_{\mu\nu} G^a_A + \Sigma^a_{\mu\nu} T^a \ .
\ee

We also need multiplier field. 
In case of $N=4$, the multiplier field has the following form 
\bea
\label{mul4}
\Phi&=&\phi^{ABCD}M_{ABCD}+\kappa^{KL\,ABC}N_{KL\,ABC} \nn
&& +\lambda^{ab}K^{ab}(\alpha)+\xi^{ab}K^{ab}(\beta)
+\zeta^{KL\,A}L_{KL\,A} + \rho H \ ,
\eea
and in case of $N=3$ TSA
\be
\label{mul3}
\Phi=\phi^{ABCD}M_{ABCD}+\kappa^{aABC}N^a_{ABC}
+\lambda^{ab}K^{ab}+\zeta^A L_A \ .
\ee

The Lagrangeans of $N=3$ and $N=4$ two-form 
supergravity have the following form
\be
\label{lag0}
{\cal L}=\epsilon^{\mu\nu\rho\sigma}\left\{
\tr R_{\mu\nu}X_{\rho\sigma}
+\Lambda \tr X_{\mu\nu}X_{\rho\sigma} 
+\tr \Phi(X_{\mu\nu}\otimes X_{\rho\sigma}) \right\}\ .
\ee
Here $\Lambda$ is a cosmological constant.
In the component fields, the terms in the 
Lagrangean of $N=4$ two-form 
supergravity are given by\footnote{
Let $Q_i$'s be the generators of $N=4$ or $N=3$ TSA, 
then the trace for the two tensor products are defined by
\[
\tr (Q_1\otimes Q_2)(Q_3 \otimes Q_4) 
\equiv (-1)^{s_{23}}\tr Q_1 Q_3 \tr Q_2 Q_4\ .
\] 
Here $s_{23}=1$ if both of $Q_2$ and $Q_3$ are 
$G^a_A$ or $G_{KL\,A}$, otherwise $s_{23}=0$.
}
\bea
\label{lag4}
\tr R_{\mu\nu}X_{\rho\sigma}
&=& {1 \over \alpha}
\{\partial_\mu A^a_\nu-\partial_\nu A^a_\mu
+i\epsilon^{abc}A^b_\mu A^c_\nu \nn
&&\ -\alpha T^a_{KM}\epsilon_{LN}\epsilon_{AB}
\psi^{KL\,A}_\mu\psi^{MN\,B}_\nu\}\Theta^a_{\rho\sigma}  \nn
&& +{1 \over \beta}
\{\partial_\mu B^a_\nu-\partial_\nu B^a_\mu
+i\epsilon^{abc}B^b_\mu B^c_\nu \nn
&&\ -\beta \epsilon_{KM}T^a_{LN}\epsilon_{AB}
\psi^{KL\,A}_\mu\psi^{MN\,B}_\nu\} \Pi^a_{\rho\sigma}  \nn
&& +\epsilon_{KP}\epsilon_{LQ}\epsilon_{AC}
\{\partial_\mu\psi^{KL\,A}_\nu
-\partial_\nu\psi^{KL\,A}_\mu \nn
&&\ +T^{a\ A}_B(\omega_\mu^a \psi_\nu^{KL\,B}
-\omega_\nu^a\psi_\mu^{KL\,B}) \nn
&&\ \ +T^{a\ K}_M(A_\mu^a \psi_\nu^{ML\,A}
-A_\nu^a\psi_\mu^{ML\,A}) \nn
&&\ \ +T^{a\ L}_N(B^a \psi_\nu^{KN\,A}
-B_\nu^a\psi_\mu^{KN\,A})\} \chi^{PQC}_{\rho\sigma} \nn
&& +\{\partial_\mu\omega_\nu^a-\partial_\nu\omega_\mu^a
+i\epsilon^{abc}\omega_\mu^b\omega_\nu^c \nn
&& -\epsilon_{KM}\epsilon_{LN}T^a_{AB}
\psi^{KL\,A}_\mu\psi^{MN\,B}_\nu\}  \Sigma^a_{\rho\sigma} \ , \\
\tr X_{\mu\nu}X_{\rho\sigma}&=&
{1 \over \alpha} \Theta^a_{\mu\nu}  \Theta^a_{\rho\sigma} 
+ {1 \over \beta}\Pi^a_{\mu\nu} \Pi^a_{\rho\sigma}  
+\epsilon_{KM}\epsilon_{LN}\epsilon_{AB}
\chi^{KL\, A}_{\mu\nu} \chi^{MN\, B}_{\rho\sigma} \nn
&&\ + \Sigma^a_{\mu\nu} \Sigma^a_{\rho\sigma} \ , \\
\tr \Phi(X_{\mu\nu}\otimes X_{\rho\sigma}) &=&
\phi^{ABCD}T^a_{(AB}T^b_{CD)}
\Sigma^a_{\mu\nu} \Sigma^b_{\rho\sigma} \nn
&& +\epsilon_{KM}\epsilon_{LN}
T^a_{(AB}\epsilon_{C)D}\kappa^{KL\,ABC}(\chi^{MN\,D}_{\mu\nu} 
\Sigma^a_{\rho\sigma} 
+ \Sigma^a_{\mu\nu} \chi^{MN\,D}_{\rho\sigma}) \nn
&& +\xi^{ab}\{\Theta^a_{\mu\nu}\Sigma^b_{\rho\sigma}
+\Sigma^b_{\mu\nu}\Theta^a_{\rho\sigma}
-2T^a_{KM}\epsilon_{LN}T^b_{AB}\chi^{KL\, A}_{\mu\nu}
\chi^{MN\, B}_{\rho\sigma} \} \nn
&& +\lambda^{ab}\{\Pi^a_{\mu\nu}\Sigma^b_{\rho\sigma}
+\Sigma^b_{\mu\nu}\Pi^a_{\rho\sigma}
-2\epsilon_{KM}T^a_{LN}T^b_{AB}
\chi^{KL\, A}_{\mu\nu}
\chi^{MN\, B}_{\rho\sigma}\} \nn
&& -\zeta^{KL\,A}\{
{2 \over 3}(\alpha-\beta)
\epsilon_{KM}\epsilon_{LN}
T^a_{AB}(\chi^{MN\, B}_{\mu\nu}\Sigma^a_{\rho\sigma}
+\Sigma^a_{\mu\nu}\chi^{MN\, B}_{\rho\sigma}) \nn
&& +2 T^a_{KM}\epsilon_{LN}\epsilon_{AB}(\chi^{MN\, B}_{\mu\nu}\Theta^a_{\rho\sigma}
+\Theta^a_{\mu\nu}\chi^{MN\, B}_{\rho\sigma}) \nn
&& -2 \epsilon_{KM} T^a_{LN}\epsilon_{AB}(\chi^{MN\, B}_{\mu\nu}\Pi^a_{\rho\sigma}
+\Pi^a_{\mu\nu}\chi^{MN\, B}_{\rho\sigma}) \} \nn
&& + \rho \{{1 \over 3}(\alpha -\beta)\Sigma^a_{\mu\nu}
\Sigma^a_{\rho\sigma}
+ \Theta^a_{\mu\nu}\Theta^a_{\rho\sigma}
- \Pi^a_{\mu\nu}\Pi^a_{\rho\sigma} \nn
&& -{1 \over 2}(\alpha -\beta)\epsilon_{KM}\epsilon_{LN}
\epsilon_{AB}\chi^{KL\,A}\chi^{MN\,B}\} \ ,
\eea
and that of $N=3$ two-form supergravity
\bea
\label{lag3}
\tr R_{\mu\nu}X_{\rho\sigma}
&=& -4\{\partial_\mu A^a_\nu-\partial_\nu A^a_\mu
+i\epsilon^{abc}A^b_\mu A^c_\nu \nn
&& -{i \over 8}i\epsilon^{abc}\epsilon_{AB}
\psi^{b\,A}_\mu\psi^{c\,B}_\nu\} \Pi^a_{\rho\sigma}  \nn
&& -{1 \over 2}\epsilon_{AC}
\{\partial_\mu\psi^{a\,A}_\nu
-\partial_\nu\psi^{a\,A}_\mu
+T^{b\ A}_B(\omega_\mu^b \psi_\nu^{a\,B}
-\omega_\nu^b\psi_\mu^{a\,B}) \nn
&&\ \ +i\epsilon^{abc}(A_\mu^b \psi_\nu^{c\,A}
-A_\nu^b\psi_\mu^{c\,A})\} \chi^{aC}_{\rho\sigma}  \nn
&&+\{\partial_\mu\omega_\nu^a-\partial_\nu\omega_\mu^a
+i\epsilon^{abc}\omega_\mu^b\omega_\nu^c \nn
&&+{1 \over 2}T^a_{AB}
\psi^{b\,A}_\mu\psi^{b\,B}_\nu\} \Sigma^a_{\rho\sigma} \ , \\
\tr X_{\mu\nu}X_{\rho\sigma}&=&
-4 \Pi^a_{\mu\nu} \Pi^a_{\rho\sigma}   
-{1 \over 2}\epsilon_{AB}
\chi^{aA}_{\mu\nu} \chi^{aB}_{\rho\sigma}  
+ \Sigma^a_{\mu\nu} \Sigma^a_{\rho\sigma} \ , \\
\tr \Phi(X_{\mu\nu}\otimes X_{\rho\sigma}) &=&
\phi^{ABCD}T^a_{(AB}T^b_{CD)}
\Sigma^a_{\mu\nu} \Sigma^b_{\rho\sigma} \nn
&& -{1 \over 2}\kappa^{aABC}T^b_{(AB}\epsilon_{C)D}
(\chi^{aD}_{\mu\nu} \Sigma^b_{\rho\sigma} 
+\Sigma^b_{\mu\nu} \chi^{aD}_{\rho\sigma}) \nn
&& +\lambda^{ab}\{-2(\Pi^a_{\mu\nu} \Sigma^b_{\rho\sigma}
+\Sigma^b_{\mu\nu} \Pi^a_{\rho\sigma})
+i\epsilon^{acd}T^b_{AB}\chi^{cA}_{\mu\nu} 
\chi^{dB}_{\rho\sigma} \} \nn
&& +2\epsilon_{AB}\zeta^A 
(\chi^{aB}_{\mu\nu} \Pi^a_{\rho\sigma} 
+\Pi^a_{\mu\nu} \chi^{aB}_{\rho\sigma} ) \ .
\eea

The integration of the multiplier field $\phi^{ABCD}$ 
gives a constraint 
\be
\label{cons1} 
\epsilon^{\mu\nu\rho\sigma}T^a_{(AB}T^b_{CD)}
\Sigma^a_{\mu\nu} \Sigma^b_{\rho\sigma} =0 \ ,
\ee
which can be solved by introducing the vierbein 
field $e^n_\nu$ 
\be
\label{sol1}
\Sigma^a_{\mu\nu} =\eta^a_{mn}e^m_\mu e^n_\nu \ .
\ee
Here $\eta^a_{mn}$ is called 
the 't Hooft symbol and defined 
by
\be
\label{thft}
\left\{
\begin{array}{l}
\eta^a_{bc}=\epsilon^{abc} \\
\eta^a_{4b}=-\eta^a_{b4}=\delta^{ab}
\end{array}
\right. \ .
\ee
The equation of motion which is obtained from the 
variation of $\omega^a_\mu$ can be solved with respect to 
$\omega^a_\mu$ algebraically, \ie $\omega^a_\mu$ 
becomes a dependent field. 

The constraints given by the integration of 
$\kappa^{KL\,ABC}$ in $N=4$ case
\be
\label{cons2}
\epsilon^{\mu\nu\rho\sigma}
T^a_{(AB}\epsilon_{C)D}(\chi^{MN\,D}_{\mu\nu} 
\Sigma^a_{\rho\sigma} 
+ \Sigma^a_{\mu\nu} \chi^{MN\,D}_{\rho\sigma}) 
=0 \ ,
\ee
and $\kappa^{aABC}$ in $N=3$ case 
\be
\label{cons3}
\epsilon^{\mu\nu\rho\sigma}T^b_{(AB}\epsilon_{C)D}
(\chi^{aD}_{\mu\nu} \Sigma^b_{\rho\sigma} 
+\Sigma^b_{\mu\nu} \chi^{aD}_{\rho\sigma}) 
=0 \ ,
\ee
are solved by introducing the right-handed 
Rarita-Schwinger field $\psi^{KL\,\dot A}_\mu$ and 
$\psi^{a\dot A}_\mu$ if we assume Eq.(\ref{sol1})
\bea
\label{rRS}
\chi^{MN\,D}_{\mu\nu}&=&
e^m_\mu \sigma^{\ A}_{m\ \dot B}
\psi^{MN\,\dot B}_\nu
-e^m_\nu \sigma^{\ A}_{m\ \dot B}
\psi^{MN\,\dot B}_\mu \ , \\
\chi^{aD}_{\mu\nu}&=&
e^m_\mu \sigma^{\ A}_{m\ \dot B}
\psi^{a\dot B}_\nu
-e^m_\nu \sigma^{\ A}_{m\ \dot B}
\psi^{a\dot B}_\mu \ . 
\eea
By counting the degrees of the freedom, we find that 
other constraints given by $\lambda^{ab}$, $\xi^{ab}$, 
$\zeta^{KL\, A}$, $\rho$ ($N=4$ case) and $\lambda^{ab}$, 
$\zeta^A$ ($N=3$ case) can be also solved. 
The constraints given by 
$\zeta^{KL\, A}$ and $\zeta^A$ tell that all of 
the $4\times 8$ ($N=4$ case) or $3\times 8$ ($N=3$ case) 
components of the right-handed Rarita-Schwinger field are
not independent but only $4\times 6$ or $3\times 6$ ones 
are independent.

The existence of the right-handed Rarita-Schwinger field 
implies that the system has the right-handed supersymmetry.
The right-handed supersymmetry appears as 
the residual symmetry of the Kalb-Ramond symmetry 
in the BF theory after imposing the constraint 
conditions \cite{nojiri}.
The Lagrangean of the BF theory has the following form
\be
\label{lagKR}
\cL_{\rm BF}=\epsilon^{\mu\nu\rho\sigma}\left\{
\tr R_{\mu\nu}X_{\rho\sigma} 
+\Lambda\tr X_{\mu\nu} X_{\rho\sigma}
\right\} \ .
\ee
The Lagrangean (\ref{lagKR}) has the large local 
symmetry which is called the Kalb-Ramond symmetry.
The transformation law of the Kalb-Ramond symmetry is given 
by 
\bea
\label{KRtr1}
\delta_{{\rm KR}} A_\mu&=& -\Lambda C_\mu \ , \nn
\delta_{{\rm KR}} X_{\mu\nu}&=&{1 \over 2}(D_\mu C_\nu-
D_\nu C_\mu) \ .
\eea
Here the parameter of the transformation $C_\mu$ is 
in the same representation of $N=3$ or $N=4$ TSA 
gauge field and 
the covariant derivative $D_\mu$ is defined by
\be
D_\mu \ \cdot\ =[\partial_\mu+V_\mu,\ \cdot \ ] \ .
\ee
Now we consider the Kalb-Ramond like transformation 
for the Lagrangean (\ref{lag0}):
\bea
\label{KRtr2}
\delta_{{\rm KR}} A_\mu&=& -\Lambda C_\mu 
- \Phi\times C_\mu \ , \nn
\delta_{{\rm KR}} X_{\mu\nu}&=&{1 \over 2}(D_\mu C_\nu-
D_\nu C_\mu) \ .
\eea
Here the product $R\times S$ of two operators 
$R=\sum_{ij} r_{ij}Q^i\otimes Q^j$ and $S=\sum_i s_i Q^i$ 
($Q_i$'s are the generators of $N=4$ or $N=3$ TSA), 
is defined by
\be
\label{pr}
R \times S \equiv \sum_{ijk} r_{jk} s_i g^{ij} Q^k \ .
\ee
Here $g^{ij}$ is defined by the trace of the two 
operators $g^{ij}\equiv \tr Q^i Q^j$ in Equation (\ref{tr4}) 
or (\ref{tr3}).
Then the change of the Lagrangean (\ref{lag0}) is given by
\be
\label{change}
\delta_{{\rm KR}} \cL = -\epsilon^{\mu\nu\rho\sigma}
\tr D_\mu  \Phi (C_\nu \otimes X_{\rho\sigma}) 
+\ \mbox{total derivative} \ .
\ee
This tells that the Lagrangean (\ref{lag0}) is invariant if 
\be
\label{KRcon}
\epsilon^{\mu\nu\rho\sigma}(
C_\nu \otimes X_{\rho\sigma})|_{\mbox{\footnotesize 
$N=3$ or $N=4$ 
spin 2 representation part}}=0 \ .
\ee
Equation (\ref{KRcon}) does not have solution in 
general, which can be found by the counting of the 
degrees of freedom. 
If we assume, however, that the two-form field $X_{\mu\nu}$ 
satisfies the constraints given by the multiplier 
field $\Phi$, there is a unique solution,\footnote{
In this sense, the ``symmetry'' discussed in the following 
is not the real symmetry. The ``symmetry'' expresses the 
redundancy when we write the two-form field $X_{\mu\nu}$ 
by the vierbein \etc after solving the constraints 
given by the multiplier field $\Phi$. 
Therefore the ``symmetry'' can be regarded as a kind of 
hidden local symmetry.
}
\be
\label{Cmu}
C_\mu=X_{\mu\nu}v^\nu\ .
\ee
Here $v^\nu$ is an arbitrary vector field parameter.
The Kalb-Ramond like transformation (\ref{KRtr2}) given 
by (\ref{Cmu})  contains the general coordinate 
transformation.
Now we consider the case where $v^\mu$ is given by 
\be
\label{rhsusy}
v^\mu = \left\{
\begin{array}{ll}
\epsilon_{KM}\epsilon_{LN}
\varepsilon^{KL\, \dot A}\sigma^\nu_{\dot A A}
\chi^{MN\,A\ \mu}_{\ \ \ \ \ \ \nu} &(N=4) \\
\varepsilon^{a \dot A}\sigma^\nu_{\dot A A}
\chi^{aA\ \mu}_{\ \ \nu} &(N=3) 
\end{array}\right.\ .
\ee
Here $\varepsilon^{KL\,\dot A}$ and $\varepsilon^{a\dot A}$ 
are right-handed spinors. 
Since the vector parameter $v^\mu$ is field dependent, 
$v^\mu$ is transformed by the left-handed supersymmetry 
transformation, which is generated by 
$\eta^{KL\,A}G_{KL\,A}$ or $\eta^{aA}G^a_A$ 
($\eta^{KL\,A}$ and $\eta^{aA}$ are left-handed spinor 
parameters), as follows
\be
\label{comm}
v^\mu \rightarrow v^\mu + \delta v^\mu \ , \hskip 1cm 
\delta v^\mu \sim \left\{
\begin{array}{ll}
\epsilon_{KM}\epsilon_{LN}
\varepsilon^{KL\, \dot A}\sigma^\mu_{\dot A A}
\eta^{MN\,A} &(N=4) \\
\varepsilon^{a \dot A}\sigma^\mu_{\dot A A}
\eta^{aA}  &(N=3) 
\end{array}\right.\ .
\ee
Therefore the commutator of the left-handed supersymmetry 
transformation and the Kalb-Ramond like transformation 
(\ref{KRtr2}) whose parameter is given by 
Eqs.(\ref{Cmu}) and (\ref{rhsusy}) contains 
the general coordinate transformation whose parameter is 
given by (\ref{comm}).
The commutator corresponds to the usual commutator 
of two supercharge:
\be
\label{qqp}
[\epsilon Q, \eta Q]=\bar \epsilon \gamma^\mu \eta P_\mu \ .
\ee
This tells that the Kalb-Ramond like transformation given by 
Eqs.(\ref{KRtr2}), (\ref{Cmu}) and (\ref{rhsusy})  can 
be regarded 
as the right-handed supersymmetry.

In summary, we have constructed the Lagrangeans 
of $N=3$ and $N=4$ two-form supergravities as gauge theories. 
The gauge algebras used here can be identified 
with the subalgebra of 
$SU(2)\times SU(2)\times U(1)$-extended $N=4$ superconformal 
algebra.
We have also shown that 
the right-handed supersymmetry appears as 
the residual symmetry of the Kalb-Ramond symmetry 
in the BF theory after imposing the constraint 
conditions.

We are indebted to A. Sugamoto for the discussion and the reading 
the manuscript.

\end{document}